\documentclass[9pt,conference]{IEEEtran}

\usepackage{amssymb,amsmath,graphicx,booktabs}
\usepackage{multirow}
\usepackage{makecell}
\usepackage{url}
\usepackage{hyperref}

\PassOptionsToPackage{hyphens}{url}\usepackage{hyperref}

\usepackage{xcolor}
\usepackage{paralist}
\usepackage{enumitem}
\usepackage[caption=false]{subfig}
\usepackage{float}

\usepackage{multicol}
\usepackage{tabularx}
\usepackage{color}
\usepackage{booktabs}
\usepackage[dvips]{epsfig}
\usepackage{setspace}
\usepackage[utf8]{inputenc} 

\newcommand{\natsubjects}{12}

\usepackage{cite}
\usepackage{amsmath,amssymb,amsfonts}
\usepackage{algorithmic}
\usepackage{graphicx}
\usepackage{textcomp}
\usepackage{xcolor}
\def\BibTeX{{\rm B\kern-.05em{\sc i\kern-.025em b}\kern-.08em
    T\kern-.1667em\lower.7ex\hbox{E}\kern-.125emX}}
\begin{document}
\title{Description-based Controllable Text-to-Speech with Cross-Lingual Voice Control}

\author{\IEEEauthorblockN{Ryuichi Yamamoto}
\IEEEauthorblockA{\textit{LY Corporation} \\
Tokyo, Japan \\
ryuichi.yamamoto@lycorp.co.jp}
\and
\IEEEauthorblockN{Yuma Shirahata}
\IEEEauthorblockA{\textit{LY Corporation} \\
Tokyo, Japan \\
yuma.shirahata@lycorp.co.jp}
\and
\IEEEauthorblockN{Masaya Kawamura}
\IEEEauthorblockA{\textit{LY Corporation} \\
Tokyo, Japan \\
kawamura.masaya@lycorp.co.jp}
\and
\IEEEauthorblockN{Kentaro Tachibana}
\IEEEauthorblockA{\textit{LY Corporation} \\
Tokyo, Japan \\
kentaro.tachibana@lycorp.co.jp}
}

\maketitle

\begin{abstract}
We propose a novel description-based controllable text-to-speech (TTS) method with cross-lingual control capability. To address the lack of audio-description paired data in the target language, we combine a TTS model trained on the target language with a description control model trained on another language, which maps input text descriptions to the conditional features of the TTS model. These two models share disentangled timbre and style representations based on self-supervised learning (SSL), allowing for disentangled voice control, such as controlling speaking styles while retaining the original timbre. Furthermore, because the SSL-based timbre and style representations are language-agnostic, combining the TTS and description control models while sharing the same embedding space effectively enables cross-lingual control of voice characteristics. Experiments on English and Japanese TTS demonstrate that our method achieves high naturalness and controllability for both languages, even though no Japanese audio-description pairs are used.
\end{abstract}


\section{Introduction}


Speech generation models trained on large-scale datasets have demonstrated not only human-level quality in synthetic speech but also a strong capability to generate diverse speaker characteristics~\cite{tan2024naturalspeech, shen2023naturalspeech, vyas2023audiobox, lajszczak2024base,chen2024vall}.
Furthermore, description\footnote{
Throughout this paper, we use the term \textit{description} instead of \textit{prompt} since the latter may be used as non-textual information~\cite{vyas2023audiobox}.
}-based controllable text-to-speech (TTS)~\cite{guo2023prompttts,yang2023instructtts,liu2023promptstyle,leng2023prompttts,zhou2024voxinstruct}, which uses natural language descriptions to intuitively control voice characteristics, has become a promising research area due to the advances of large language models (LLMs)~\cite{ouyang2022training,touvron2023llama}.



Despite the success of description-based TTS, the need for labeled datasets remains a significant challenge. Specifically, it is difficult to collect annotations for audio-description pairs, although these are necessary to learn the mapping between text descriptions and the target audio. Since the quality of the descriptions is crucial for description-based TTS, some prior works have constructed dedicated datasets with audio-description pairs through manual annotations, such as PromptSpeech~\cite{guo2023prompttts}, NLSpeech~\cite{yang2023instructtts}, Coco-Nut~\cite{watanabe2023coco}, TextrolSpeech~\cite{ji2023textrolspeech}, and LibriTTS-P~\cite{kawamura24_interspeech}. Although these datasets are valuable, they are limited to a few popular languages. This data scarcity greatly limits applications and further research in other languages.

\begin{figure*}[t]
  \centerline{\epsfig{figure=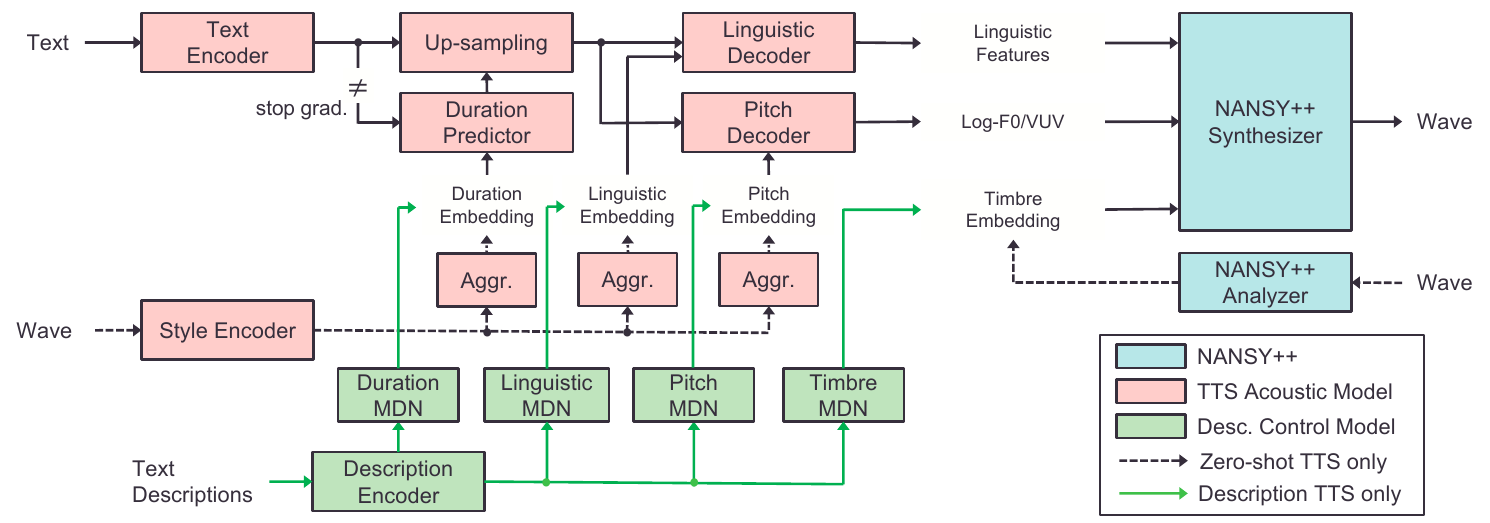,width=0.92\linewidth}}
  \caption{Overview of our proposed TTS framework. A NANSY++ model (blue) is used as the backbone model to obtain disentangled representations and convert them back to the waveform. The TTS acoustic model (red) converts text to NANSY++ features. It utilizes a style encoder and feature aggregators (aggr.) to extract style embeddings. The description control model (green) predicts style and timbre embeddings from the input text descriptions. 
  }
  \label{fig:overview}
  \vspace{-3mm}
\end{figure*}

To address the aforementioned issue, we propose a novel method that enables cross-lingual~\footnote{In this work, cross-lingual controllability refers to the ability to control for a language that is different from the one on which the description control model was trained.} description-based control without relying on audio-description pairs in the target language. 
The key idea is to combine a TTS model trained on the target language with a description control model trained on another language, sharing the same timbre and style representations to enable cross-lingual controllability. The description control model converts input text descriptions (e.g., ``\textit{a man speaks slowly with a soft voice}'') into timbre and style embeddings, which are used as the conditional features for the TTS model.
As depicted in Figure~\ref{fig:overview}, our TTS framework consists of three components: (1) a neural analysis and synthesis (NANSY++~\cite{choi2022nansy++}) model, a TTS acoustic model, (3) and a description control model.
Table~\ref{tab:data_requirements} summarizes the data requirements for each model.
Thanks to the modularized design, these models can be separately trained on different datasets.

\begin{table}[t]
\centering
\caption{Data requirements for the major components of our description-based controllable TTS framework.}
\label{tab:data_requirements}
\vspace{-2mm}
\scalebox{0.92}{
\begin{tabular}{lccc}
\hline
 & \multirow{2}{*}{NANSY++} & TTS Acoustic & Desc. Control \\
 &  & Model & Model \\ \hline
Audio & \checkmark & \checkmark & \checkmark \\
Transcriptions &  & \checkmark &  \\
Descriptions &  &  & \checkmark \\ \hline
\end{tabular}
}
\vspace{-4mm}
\end{table}


In detail, we first train a NANSY++ backbone model that learns disentangled speech representations based on self-supervised learning (SSL) with unlabeled audio-only datasets. 
Then, we train a TTS acoustic model that converts text into NANSY++'s disentangled features with the target language's paired dataset of transcription and audio.
In the TTS acoustic model, we utilize an SSL-based style encoder to extract style embeddings from input waveforms~\cite{fujita2023zero}.
Finally, we train a description control model that maps the text descriptions to the timbre and style embeddings with an audio-description paired dataset in a different language.
Notably, NANSY++'s disentanglement capability allows for disentangled voice control, such as controlling speaking style while retaining the original timbre.
Furthermore, because the SSL-based timbre and style representations can be regarded as language-agnostic~\cite{nandan2020language,li2023quantitative}, combining the TTS model and the description control model while sharing the same embedding space effectively enables control of voice characteristics across different languages.

We perform English and Japanese TTS experiments for evaluations.
The listening test results for naturalness and audio-description consistency show that our method achieves high naturalness and controllability for both languages, even if no Japanese audio-description pairs are used for training.
Furthermore, additional objective evaluations demonstrate that our method allows for much better fine-grained and disentangled pitch and speaking speed control than the baseline systems.
Audio samples are available on our demo page~\footnote{
\label{demo}\url{https://r9y9.github.io/projects/nansyttspp/}
}.

\section{Method}
\label{sec:method}

\subsection{Neural analysis and synthesis}
\label{ssec:anasyn}

As shown in Figure~\ref{fig:overview}, 
we employ a NANSY++~\cite{choi2022nansy++} backbone model to learn disentangled speech representations: linguistic features, fundamental frequency ($F_0$), and timbre embedding, which are used as intermediate features of our TTS model. 
In practice, $F_0$ is converted to continuous log-$F_0$ and voiced/unvoiced flags (V/UV)~\cite{yu2010continuous}.
The NANSY++ model consists of two modules: analyzer and synthesizer. 
The analyzer decomposes a speech waveform into the disentangled representations while the synthesizer converts them back to the waveform.
The NANSY++ model utilizes contrastive learning with information perturbation~\cite{choi2021neural}, effectively performing disentanglement in a self-supervised manner.

For simplicity, we make several changes to NANSY++: (1) we replace an SSL $F_0$ estimator with a signal processing method~\cite{song2017effective}, and (2) we use a reference encoder~\cite{skerry2018towards} to extract a global timbre embedding instead of a time-varying timbre embedding.
We also remove the use of global style tokens~\cite{wang2018style} since the interpretability of the learned tokens is beyond the scope of our research.

\subsection{Acoustic model}
\label{ssec:acousitc}

Our TTS acoustic model predicts the NANSY++'s disentangled features from input text.
The acoustic model adopts a similar architecture to that of NANSY-TTS~\cite{choi2022nansy++}, capable of zero-shot TTS where a short audio clip is used to clone the voice characteristics.
The model comprises a text encoder, duration predictor, style encoder, linguistic decoder, and pitch decoder. We upsample the text-level features to frame-level by estimated phoneme durations. 

The style encoder is the critical component for our controllable TTS. 
In particular, it learns to extract style embeddings from input speech for predicting duration, linguistic features, and pitch.
We denote the corresponding learned vector representations as duration embedding, linguistic embedding, and pitch embedding, respectively.
Specifically, we use wav2vec 2.0~\cite{baevski2020wav2vec} as a style encoder to extract frame-level hidden features. 
Then, we use feature aggregators that compress the frame-level features to fixed-dimensional embeddings~\cite{fujita2023zero}. 
For feature aggregation, we apply a weighted sum of hidden activations for each layer of Transformer~\cite{vaswani2017attention}, followed by bi-directional long short-term memory (LSTM)~\cite{hochreiter1997long} and feed-forward attention~\cite{raffel2015feed}. 
Note that simpler mean pooling works well, but LSTM and attention worked better.

\subsection{Description control model}
\label{ssec:desc}

The description control model predicts the timbre and style embeddings from the input text description.
The model consists of a description encoder and four mixture density networks (MDNs)~\cite{Bishop94mixturedensity}.
The description encoder converts the text descriptions to hidden features, and the MDNs model the conditional probability distributions of four embeddings.
Note that the timbre embedding is the intermediate feature of NANSY++, which is assumed to be independent of speaking style. In contrast, the other embeddings are the style-related features of the TTS acoustic model. 
The model architecture is the same as that of PromptTTS++~\cite{shimizu2023prompttts++}, except that we use separate MDNs for modeling each embedding as the effectiveness of the separate modeling was confirmed in the prior work~\cite{fujita2023zero}. 

\subsection{Training}

Due to the modularized structure of our framework, we adopt the following three-stage training.

\noindent\textbf{NANSY++:}
First, we train a NANSY++ model using an unlabeled audio-only dataset, which may contain diverse data from multiple languages. 
In this stage, contrastive loss, multi-resolution short-time Fourier transform (STFT) loss, and adversarial loss are employed to obtain disentangled representations and enable high-fidelity reconstruction of waveforms.

\noindent\textbf{TTS acoustic model:}
Second, we train the TTS acoustic model using the target language's TTS dataset.
To extract the target features of the TTS acoustic model, we use a pre-trained NANSY++ analyzer.
We use L1 losses between predicted and target NANSY++ features as the optimization criteria. 
In addition, we use log-domain L1 loss between predicted and ground-truth phoneme durations~\cite{ren2020fastspeech}.

\noindent\textbf{Description control model:}
Lastly, we train the description control model using the audio-description pairs of another language. 
We use the pre-trained NANSY++ analyzer and TTS acoustic model to extract timbre and style embeddings, which are used as the target features.
We use negative log-likelihood losses to train MDNs.

\subsection{Inference}

To perform controllable TTS with text descriptions, the description control model converts the text descriptions to style and timbre embeddings. Then, by injecting the embeddings into the TTS acoustic model and the NANSY++ synthesizer, we can obtain the output speech. On the other hand, for zero-shot TTS, we pass a reference waveform to the style encoder and NANSY++ analyzer, and the extracted embeddings can then be used to synthesize the output waveform.
Note that cross-lingual voice control can be performed by replacing the TTS acoustic model for another language while sharing the description control model and NANSY++.


\section{Experimental evaluations}
\label{sec:exp}

We perform English and Japanese TTS experiments, where text descriptions are unavailable for the Japanese dataset. 


\subsection{Data}

\noindent\textbf{Unlabeled audio-only dataset}:
To train the NANSY++ backbone model, we used 1,281 hours of speech data, which includes publicly available LibriTTS-R~\cite{koizumi2023libritts} and our internal Japanese dataset. 
The internal dataset contains studio-quality recordings with various speaking styles and emotions, such as happy, sad, and angry. 
The total number of speakers was 4,158.
All audio data was sampled at 24~kHz.

\noindent\textbf{TTS dataset}:
We selected the subset of the internal Japanese corpus as the TTS dataset. 
In detail, 180~K utterances were used for training. 
For validation and test sets, we used 5850 utterances each.
The number of speakers is seventeen, and the total amount of data is 208 hours.
We also used LibriTTS-R as the English TTS dataset.
The phoneme durations were manually labeled for the Japanese dataset and estimated by the Montreal forced aligner~\cite{mcauliffe2017montreal} for LibriTTS-R.

\noindent\textbf{Audio-description paired dataset}:
We used LibriTTS-R and LibriTTS-P~\footnote{
\url{https://github.com/line/LibriTTS-P}
} as the English audio-description paired dataset.
The dataset contains (1) manual annotations on human perceptions of speaker characteristics and (2) synthetic annotations on speaking style, such as pitch, duration, and energy.
We followed the same method as LibriTTS-P~\cite{kawamura24_interspeech} to construct the text descriptions for each utterance.
We separated ten speakers for evaluation, specifically with the speaker IDs: 121, 237, 260, 908, 1089, 1188, 1284, 1580, 1995, and 2300. The rest of the dataset was split into training and validation sets, with the split based on 2\% of the speakers for validation.


\subsection{Models}


\noindent\textbf{NANSY++}:
As part of the NANSY++ analyzer, we used a reference encoder to extract timbre embedding from 80-dimensional log-scale mel-spectrograms.
We used six convolutional layers with output channels set to 128, 128, 256, 256, 512, and 512, respectively. 
The number of dimensions of the timbre embedding was set to 192. 
To extract linguistic contents, we used a pre-trained ContentVec~\footnote{\url{https://github.com/auspicious3000/contentvec}}~\cite{qian2022contentvec}.
A linguistic encoder converted the 768-dimensional ContentVec features to 128-dimensional linguistic features.
For the NANSY++ synthesizer, we adopted the same model architecture as Period~VITS's decoder~\cite{shirahata2023period}. 
The input of the synthesizer is composed of 194-dimensional features, including 192-dimensional hidden features of a frame-level synthesizer, continuous log-$F_0$, and V/UV.
The other details followed the original NANSY++~\cite{choi2022nansy++}.

\noindent\textbf{TTS acoustic model}:
We used a Conformer encoder~\cite{gulati2020conformer} as the text encoder. 
The number of layers, attention heads, attention dimension, and dimension of the feed-forward module were set to 4, 4, 192, and 784, respectively.
As the style encoder, we used a pre-trained wav2vec 2.0\footnote{\url{https://huggingface.co/facebook/mms-300m}} trained on massive multilingual datasets containing over 1,400 languages.
Bidirectional LSTM aggregated the 1024-dimensional hidden features of wav2vec 2.0 to 192-dimensional style embeddings followed by a feed-forward attention.
The other details followed the original NANSY-TTS~\cite{choi2022nansy++}.

\noindent\textbf{Description control model}:
We followed the same configurations as PromptTTS++~\cite{shimizu2023prompttts++}.
As the description encoder, we used a pre-trained BERT~\footnote{\url{https://huggingface.co/bert-base-uncased}}. 
During training, we fixed the parameters of BERT except for the last attention layer~\cite{liu2023promptstyle}.
The number of mixtures was set to 10 for all MDNs.



Table~\ref{tab:mos_test} summarizes the TTS systems investigated in our experiments.
To investigate the effectiveness of the timbre-style disentangled representations, we used a baseline system that resembles PromptTTS++~\cite{shimizu2023prompttts++} with 80-dimensional log-scale mel-spectrogram as intermediate features.
We used a universal neural vocoder for the baseline system that generates speech from a log-scale mel-spectrogram. 
For a fair comparison, the vocoder architecture was the same as that of the NANSY++ synthesizer.
The vocoder was trained on the same dataset as NANSY++ for 1800~K steps. 
The TTS acoustic model used a denoising diffusion probabilistic model~\cite{ho2020denoising}. 
Two separate MDNs were used to model style embeddings for the duration and diffusion-based acoustic models, respectively.
To confirm the effectiveness of the SSL-based style encoder, we investigated the systems with a reference encoder for both the baseline and proposed methods. We also investigated a zero-shot TTS system for comparison, where the style and timbre embeddings were extracted from the reference speech.

For training all the models, we used AdamW optimizer~\cite{loshchilov2017decoupled}.
We used an exponential decay learning rate scheduler. 
The initial learning rate was 0.002.
The decay coefficient for each epoch was 0.96 for NANSY++ and 0.9975 for the other models.
We trained the NANSY++ model, the TTS acoustic model, and the description control model for 1800~K, 200~K, and 200~K steps, respectively.

\noindent\textbf{Monolingual vs. cross-lingual}:
For the monolingual baseline system, we only trained the TTS acoustic and description control models for English. No Japanese monolingual system was used because audio-description pairs were unavailable for Japanese. We trained two TTS acoustic models for each cross-lingual system for Japanese and English while sharing a description control model and feature aggregators across two languages. The shared description control model was trained solely with the English audio-description pairs.

\subsection{Naturalness and consistency evaluations}

\begin{table*}[!t]
\begin{center}         
\caption{
Naturalness and audio-description consistency MOS test results for English and Japanese TTS. 
ML-, CL-, and ZS- represent monolingual, cross-lingual, and zero-shot TTS systems, respectively.
Bold font denotes the best score among all description-based TTS systems.
}
\label{tab:mos_test}
\scalebox{0.93}{
\begin{tabular}{l|l|l|c|c|c|c}
\toprule
 &  &  & \multicolumn{2}{c|}{English} & \multicolumn{2}{c}{Japanese} \\
System & Model Name & Style Encoder & MOS (nat) & MOS (cons) & MOS (nat) & MOS (cons) \\
\midrule
M1 & ML-PromptTTS++ & Reference Encoder & $3.64 \pm 0.11$ & $3.27 \pm 0.08$ & - & - \\
C1 & CL-PromptTTS++ & Reference Encoder & $3.40 \pm 0.12$ & $3.15 \pm 0.08$ & $3.11 \pm 0.11$ & $2.74 \pm 0.11$ \\
C2 & CL-PromptTTS++-w2v & wav2vec 2.0 & $3.38 \pm 0.12$ & $3.17 \pm 0.08$ & $2.15 \pm 0.09$ & $2.88 \pm 0.10$ \\
C3 & CL-NANSY-TTS & Reference Encoder & $3.73 \pm 0.10$ & $3.20 \pm 0.08$ & $\mathbf{4.46 \pm 0.08}$ & $3.13 \pm 0.09$ \\
C4 & CL-NANSY-TTS-w2v (proposed) & wav2vec 2.0 & $\mathbf{4.21 \pm 0.09}$ & $\mathbf{3.43 \pm 0.07}$ & $3.99 \pm 0.09$ & $\mathbf{3.34 \pm 0.08}$ \\
\midrule
Z1 & ZS-NANSY-TTS-w2v & wav2vec 2.0 & $3.99 \pm 0.11$ & $3.34 \pm 0.08$ & $3.68 \pm 0.10$ & $3.37 \pm 0.08$ \\
\midrule
R1 & Reference & - & $4.58 \pm 0.07$ & $3.39 \pm 0.08$ & - & - \\
\bottomrule
\end{tabular}
}
\end{center}
\vspace{-2mm}
\end{table*}

\begin{table*}[!t]
\begin{center}
\caption{
Style controllability evaluation results for English and Japanese TTS.
All correlation and similarity numbers were calculated as the average over the test set.
Higher is better for all metrics.
}
\label{tab:controllability_eval}
\scalebox{0.93}{
\begin{tabular}{l|l|l|cccc|cccc}
\toprule
 & & & \multicolumn{4}{c|}{English} & \multicolumn{4}{c}{Japanese} \\
\cmidrule(lr){4-7} \cmidrule(lr){8-11}
System & Model Name & Style Encoder & P-Corr & P-SIM & S-Corr & S-SIM & P-Corr & P-SIM & S-Corr & S-SIM \\
\midrule
M1 & ML-PromptTTS++ & Reference Encoder & 0.872 & 0.917 & 0.285 & 0.914 & - & - & - & - \\
C1 & CL-PromptTTS++ & Reference Encoder & 0.896 & 0.883 & 0.151 & 0.896 & 0.899 & 0.923 & 0.185 & 0.947 \\
C2 & CL-PromptTTS++-w2v & wav2vec 2.0 & $\mathbf{0.933}$ & 0.890 & 0.461 & 0.914 & $\mathbf{0.945}$ & 0.910 & 0.866 & 0.951 \\
C3 & CL-NANSY-TTS & Reference Encoder & 0.887 & $\mathbf{0.948}$ & 0.025 & $0.956$ & 0.899 & $\mathbf{0.960}$ & 0.085 & 0.969 \\
C4 & CL-NANSY-TTS-w2v (proposed) & wav2vec 2.0 & 0.930 & 0.938 & $\mathbf{0.647}$ & $\mathbf{0.962}$ & 0.937 & 0.950 & $\mathbf{0.920}$ & $\mathbf{0.975}$ \\
\bottomrule
\end{tabular}
}
\end{center}
\vspace{-3mm}
\end{table*}

We performed two subjective listening tests for English and Japanese TTS: a 5-point naturalness mean opinion score (MOS) and a 4-point audio-description consistency MOS test. 
For the former, human raters are asked to make quality judgments about the speech samples using the following five possible responses: 1 = Bad; 2 = Poor; 3 =
Fair; 4 = Good; and 5 = Excellent. 
For the latter, given a pair of a speech sample and corresponding text description, raters are asked to judge the audio-description consistency with the following choices: 1 = Inconsistent; 2 = Somewhat inconsistent; 3 = Somewhat consistent; and 4 = Consistent. 
The number of subjects for each MOS test was \natsubjects.
For English TTS, we selected three random utterances for each test speaker of LibriTTS-R.
Then, using the corresponding descriptions and transcriptions as input, we generated the synthetic speech samples.
For Japanese TTS, we synthesized samples using Japanese transcriptions from our test set and descriptions from LibriTTS-R.
In total, we evaluated 30 utterances for each system.


Table~\ref{tab:mos_test} shows the evaluation results. 
The results can be summarized as follows:
\begin{inparaenum}[(1)]
\item The use of timbre-style disentangled features was effective for cross-lingual controllability; even though the monolingual baseline system (M1) performed well in both naturalness and audio-description consistency, the cross-lingual baseline systems with the entangled features (i.e., log-melspectrogram) performed poorly in consistency, especially for Japanese (C1 and C2).
\item Using wav2vec 2.0 as the style encoder improved audio-description consistency for both languages (C1 vs. C2 and C3 vs. C4). Note that we often encountered training instability with wav2vec 2.0 in the baseline systems, making the baseline system (C2) perform worse in naturalness.
\item Our proposed method performed best regarding audio-description consistency for both languages, significantly outperforming the baseline systems (C4~vs.~C2). Furthermore, there was no statistically significant difference (in a student’s t-test with a 5~\% significance level) in the consistency compared to zero-shot TTS and reference speech (C4~vs.~Z1 and C4~vs.~R1), suggesting sufficient controllability of our method.
\end{inparaenum}

Meanwhile, in terms of naturalness, we observed that zero-shot TTS performed much worse than the best description TTS system (C4~vs.~Z1). In addition, for Japanese TTS, the proposed method performed worse when wav2vec 2.0 was used as the style encoder (C3~vs.~C4). 
We analyzed the synthetic samples and found that those with low scores contained buzzy sounds in the unvoiced regions.
The buzzy sounds appeared perceptually noisier when the speaking speed was controlled, and it happened more with the SSL-based style encoder because of its ability to extract features relevant to duration prediction~\cite{fujita2023zero}.
We also noticed that some Japanese samples had slight foreign accents. Addressing these issues will be one of our future tasks.
We encourage readers to listen to the audio samples at our demo page~\footnotemark[3].

\subsection{Style controllability evaluations}

We performed objective evaluations on the pitch and speaking speed control to evaluate fine-grained controllability~\footnote{
We leave the energy controllability experiments for future work, as our dataset currently does not have enough variations on energy.
}.
For the test set, we generated synthetic speech with five-level pitch and speed: 1 = very low/slow; 2 = low/slow; 3 = normal; 4 = high/fast; and 5 = very high/fast by changing the input descriptions.
Then, we computed Pearson correlation (P-Corr and S-Corr) between five-level scores and mean $F_0$ and speaking speed of synthetic speech, respectively. 
We measured the number of syllables (or moras for Japanese) per second as the speaking speed.
In addition, we measured an average cosine similarity between (1) the speaker embedding extracted from the sample with normal pitch/speed and (2) the speaker embeddings extracted from the other four samples (P-SIM and S-SIM). 
These similarity scores represent how the speaker information was affected during the pitch and speed control; the higher the score, the better the disentangled style control. 
We used a pre-trained WavLM-based speaker verification model~\footnote{
\url{https://huggingface.co/microsoft/wavlm-base-plus-sv}
} for extracting speaker embeddings~\cite{chen2022wavlm}.

Table~\ref{tab:controllability_eval} shows the evaluation results, of which findings are summarized as follows: 
\begin{inparaenum}[(1)]
\item Using the wav2vec 2.0 as the style encoder significantly improved the speed controllability. 
\item The proposed method performed better in disentangled style control. Although the cross-lingual baseline system showed reasonable controllability on both pitch and speaking speed, the similarity scores were lower than the proposed method (C1 vs. C3 and C2 vs. C4), indicating the effectiveness of disentangled representations.
\item The speaking speed controllability was much better for the Japanese. This is because our Japanese dataset contained much more diverse speaking styles with different speeds. The results imply the importance of the data coverage of the training data for controllable TTS.
\end{inparaenum}

\section{Conclusion}
\label{sec:conclusion}

We proposed a description-based controllable TTS method with cross-lingual control capability. 
We demonstrated the feasibility of description-based control across different languages by employing language-agnostic disentangled representations derived from SSL-based models. Experiments on English and Japanese TTS confirmed that our method effectively enabled high naturalness and controllability, even without using audio-description pairs for Japanese data.
Future work includes exploring larger-scale datasets to enhance the generalization capability and cross-lingual controllability.

\bibliographystyle{IEEEtran}
\bibliography{tts}

\end{document}